\documentclass[%
 reprint,
 amsmath,amssymb,
 aps,prl
]{revtex4-2}

\usepackage[version=3]{mhchem} 
\usepackage[T1]{fontenc}       
\usepackage{dcolumn}
\usepackage{bm}
\usepackage{dcolumn}
\usepackage{amsmath,amsfonts,amssymb}
\usepackage{graphicx}
\usepackage{setspace}
\usepackage{tocloft}
\usepackage{bm}
\usepackage{lineno}
\usepackage{hyperref}
\allowdisplaybreaks

\usepackage{multirow}
\usepackage[normalem]{ulem}
\usepackage[table]{xcolor}

\begin{document}

\preprint{APS}

\title{Characterize localization length of disordered lattices via critical coupling effect}

\author{Fuhao Ji$^{1,+}$}
\author{Xiangqi Huang$^{1,+}$}
\author{Luxing Chen$^{1,+}$}
\author{Yuxiang Tian$^{1}$}
\author{Wenjing Li$^{1}$}
\author{Yinying Peng$^{1}$}
\author{Yuge Qiu$^{1}$}
\author{Lu Zhang$^2$}
\author{Liwei Zhang$^1$}
\author{Mingfang Yi$^{1}$}
\author{Peilong Hong$^{1,3}$}
\email{p.l.hong@aqnu.edu.cn}
\affiliation{$^1$School of Physics and Astronomy, Anqing Normal University, Anqing, Anhui 246133, China \\
$^2$School of Physics and Electronic Engineering, Jining University, Qufu, Shandong, 273155, China\\
$^3$The MOE Key Laboratory of Weak-Light Nonlinear Photonics, School of Physics and TEDA Applied Physics Institute, Nankai University, Tianjin 300071, China}
\collaboration{$^+$These authors contributed equally.}

\date{\today}

\begin{abstract}
Light localization by scattering is a fundamental mechanism driving phase transitions of wave transport in disordered systems. Characterizing the localization length in scattering systems is crucial yet challenging. In this Letter, we demonstrate a spatially matched coupling scheme using wavefront shaping to resolve the intrinsic localization length in two-dimensional disordered lattices. By tailoring the incident wavefront, our method facilitates efficient coupling of light to the minimum localized mode. We apply this approach to measure two different self-assembled lattices, and report the first observation of the critical coupling effect, which allows for the direct determination of the characteristic size of minimum localized mode. Our results reveal that for a fixed lattice periodicity, increasing the air-hole diameter significantly reduces this intrinsic localization length. This far-field metrology offers a robust framework for probing wave localization in complex media, which should be useful in various applications such as random lasing and nonlinear optics.
\end{abstract}

\maketitle

Wave transport in disordered lattices is a ubiquitous phenomenon spanning electronic, acoustic, and photonic systems~\cite{akkermans2007mesoscopic}. Multiple scattering within these disordered structures gives rise to a wealth of exotic physics, most notably Anderson localization and its associated phase transitions~\cite{lagendijk2009fifty}. Despite its fundamental importance, achieving Anderson localization in three-dimensional (3D) photonic systems remains experimentally formidable, as it typically demands a strong disorder that is difficult to realize in practice~\cite{lagendijk2009fifty,yamilov2023anderson,yamilov2025anderson}. In contrast, the scaling theory of localization predicts that even infinitesimal disorder can spontaneously induce localized states in lower-dimensional (1D or 2D) lattices~\cite{abrahams1979scaling}. Hence, the low-dimensional disordered lattices provide a promising platform for investigating the underlying physics of Anderson localization and its emerging applications in wave functionalization~\cite{garanovich2012light,segev2013anderson}.

The realization of Anderson localization in low-dimensional disordered structures has catalyzed extensive research, beginning with the prediction of transverse Anderson localization in photonic lattices~\cite{de1989transverse} and its subsequent experimental validations such as the observation of transverse localization~\cite{schwartz2007transport} and propagation-invariant Anderson speckles~\cite{kondakci2015discrete}. In recent years, the advance of wavefront shaping has unlocked the ability to selectively excite specific scattering mode~\cite{rotter2017light,cao2022shaping}, providing new possibilities for exploring scattering physics. For instance, leveraging Anderson-localized channels has enabled the enhancement of focusing of scattered light in disordered fibers~\cite{leonetti2014light}, and the targeted deposition of energy within disordered lattices is feasible by measuring the deposition matrix~\cite{bender2022depth}. Furthermore, by selectively exciting different eigenchannels in photonic lattices, the scattered light exhibits a localization-delocalization transition, highlighting the rich modal diversity inherent in these systems~\cite{hong2026universal}. Beyond fundamental physics, the ease of fabricating low-dimensional lattices has spurred diverse applications, ranging from high-efficiency random lasing~\cite{liu2014random,wei2020harnessing,sapienza2022controlling} and robust transport control~\cite{tzortzakakis2020shape,horodynski2022anti,mcintosh2024delivering} to high-capacity optical information processing~\cite{yu2024high,lib2022quantum}.


\begin{figure}[b]
	\centering
	\includegraphics[width=0.48\textwidth]{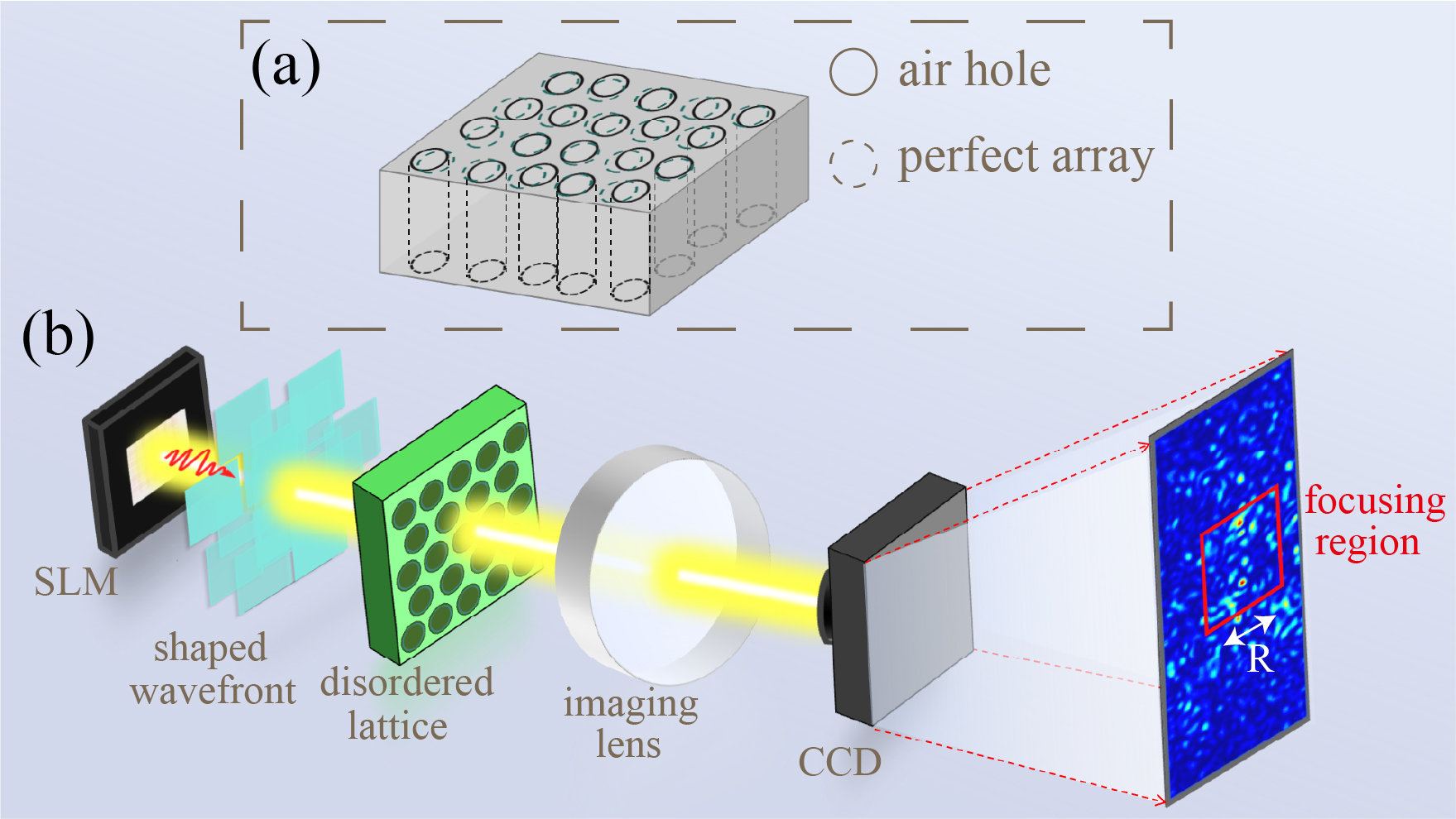}
	\caption{Scheme for characterizing localization length in a disordered lattice. (a) Schematic of the disordered photonic lattice. (b) Conceptual diagram of the spatially matched coupling technique with wavefront shaping. \label{Figure_1}}
\end{figure}


The localization size is a key parameter to understand and ultimately tailor various transport behavior in disordered systems~\cite{akkermans2007mesoscopic}.
Currently, the most direct approach involves in-plane excitation followed by the detection of out-of-plane scattered light to map the internal field distribution~\cite{spasenovic2012measuring}. However, absent a \emph{priori} knowledge of the underlying eigenmodes, an arbitrary incident wave typically excites a complex superposition of states except at the strict single-mode limit~\cite{spasenovic2012measuring,pena2014single}. An alternative strategy employs out-of-plane excitation to infer the localization length from the transverse field distribution~\cite{schwartz2007transport,levi2011disorder}. Yet, without active wavefront control, such scheme inevitably excites a superposition of localized modes with disparate sizes. The resulting interference causes the measured localization length fluctuating case by case, thereby obscuring an accurate characterization of the intrinsic localization length.

In this Letter, we overcome the challenge by implementing an out-of-plane excitation combined by adaptive wavefront shaping to achieve efficient excitation of localized state. By utilizing a spatially matched coupling technique, we report the first observation of the critical coupling effect, leading us to determine the characteristic size of the intrinsic minimum localized modes. Our work thus establishes a robust metrology for determining Anderson-localization lengths in complex media. These findings offer a versatile approach for exploring the fundamental physics of disordered systems and hold significant potential for advanced applications such as random lasing and nonlinear functional devices.

Two-dimensional photonic lattices are illustrated in Fig.~\ref{Figure_1}(a). For the out-of-plane excitation of such lattices, the propagation of the light field satisfies the wave equation~\cite{meade2008photonic}
\begin{equation}
-\frac{\partial^{2}}{\partial z^{2}}\bm{E}=H\bm{E},
\label{eq:diff}
\end{equation}
where $\bm{E}$ represents the complex amplitude of the optical field. The operator $H$ is defined as $H = \frac{\partial^{2}}{\partial x^{2}}+\frac{\partial^{2}}{\partial y^{2}}+n(x,y)k_{0}^{2}$, where $n(x,y)$ is the refractive index and $k_{0}$ is the wavenumber. Under paraxial approximation, this equation evolves to a Schrödinger equation with the $z$-axis serving as the equivalent time axis~\cite{de1989transverse}.
Consequently, the general solution of Eq.~(\ref{eq:diff}) can be expressed as $\bm{E}=\bm{A}e^{ik_{z}z}$, where $k_{z}$ corresponds to the equivalent energy. An uncontrolled incident wavefront generally excites multiple eigenmodes, while these eigenmodes have different localization lengths. Therefore, conventional out-of-plane excitation usually leads to unspecified localization sizes.

To overcome the limit, we combine out-of-plane excitation with wavefront shaping, as schematically shown in Fig.~\ref{Figure_1}(b). In theory, light scattering through the system can be modeled by a transmission matrix $\bm T$, which describes the relation between the optical fields at the input and output facets of the photonic lattice. Specifically, the incident light field is represented as a column vector $\mathbf{E}_{\text{in}}$ containing $N$ complex amplitudes, while the transmitted light field is represented as a column vector $\mathbf{E}_{\text{out}}$ containing $M$ complex amplitudes. Their relationship can then be expressed by the following equation~\cite{rotter2017light,popoff2010measuring,vellekoop2007focusing}
\begin{equation}
\mathbf{E}_{\text{out}} = \bm{T} \, \mathbf{E}_{\text{in}},
\label{eq:transmission_matrix}
\end{equation}
where $\bm T$ is an $M \times N$ transmission matrix. Its element $t_{mn}$ is a complex number representing the contribution from the $n$-th input mode to the $m$-th output mode.


\begin{figure}[b]
	\centering
	\includegraphics[width=0.48\textwidth]{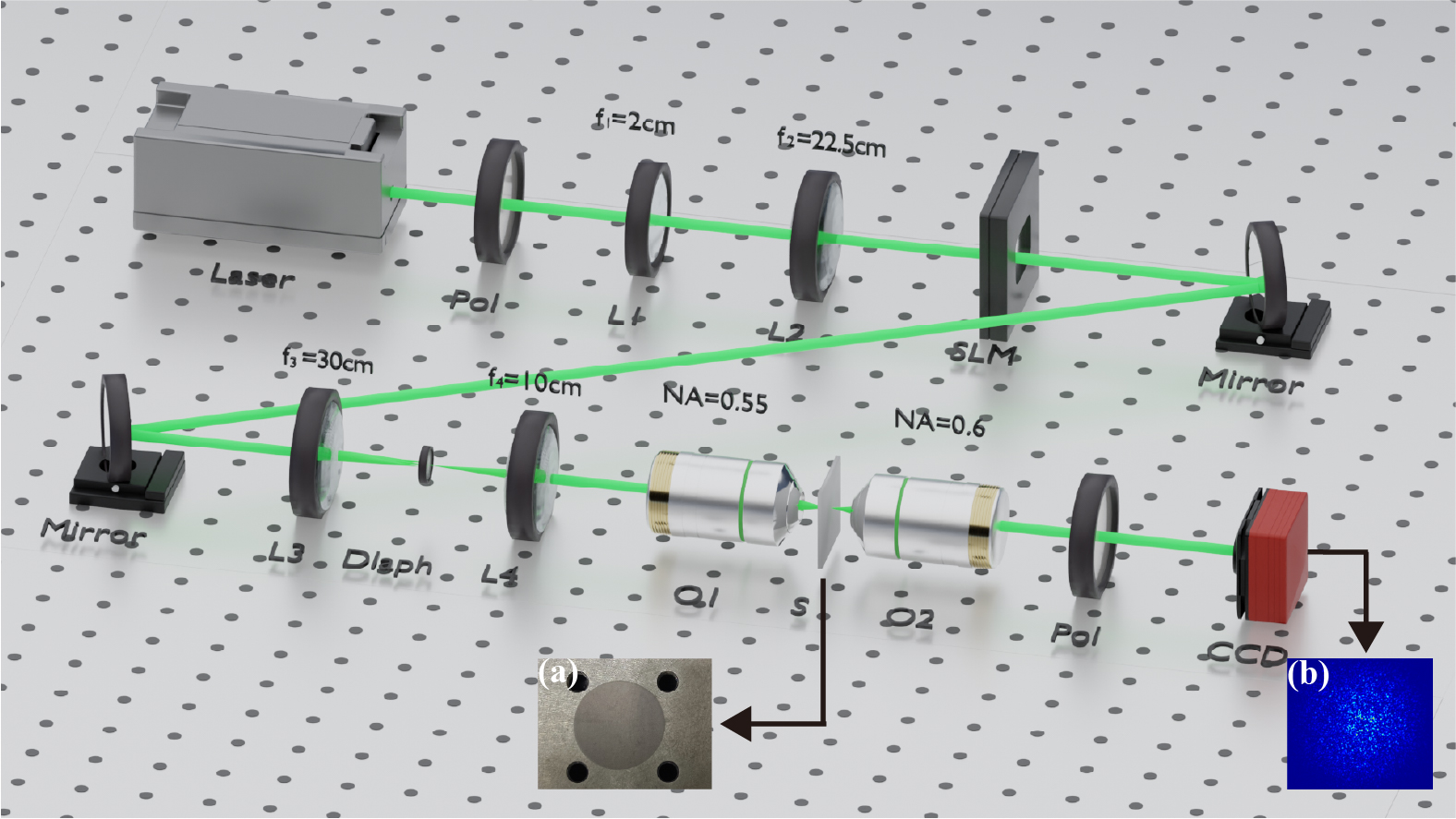}
	\caption{Schematic of the experimental setup. The inset (a) shows a photograph of the disordered sample, and (b) shows the transmitted speckle pattern without wavefront control. \label{Figure_2}}
\end{figure}

To actively control light coupling into the lattice, the most direct approach is to focus the scattered light at a target position on the transmission side. This can be achieved by shaping the complex amplitude of the incident wavefront as \( \mathbf{E}_{\text{in}} = \bm{T}^{+} \mathbf{E}_{\text{tar}} \), where \( \mathbf{E}_{\text{tar}} \) is the column vector representing the desired focused field, and \( ^{+} \) denotes the transposed conjugate of \( \bm{T} \). In practice, due to limitations such as finite controlled degrees and experimental noise, the actual transmitted field often deviates from the target $ \mathbf{E}_{\text{tar}} $~\cite{hong2024robust}. To quantify the focusing efficiency, we define the focusing fidelity \( \eta \), given by~\cite{vellekoop2008universal,hong2018three}
\begin{equation}
\eta = \frac{I_{\text{spot}}}{I_{\text{rand}}},
\label{eq:eta}
\end{equation}
where $I_{\text{spot}}$ is the intensity at the focus, and $I_{\text{rand}}$ is the total transmitted intensity corresponding to random incident light prior to optimization.
In the case of single-spot focusing, Leonetti \textit{et al.} found that the focus intensity is directly related to the excitation of localized states~\cite{leonetti2014light}, i.e. a higher degree of localization leads to stronger focusing. Consequently, in the case of coexistence of multiple eigenmodes with distinct localization lengths, a shorter localization length results in a higher energy concentration into a single spatial spot, thereby yielding a greater focusing fidelity $\eta$.

To quantify the efficiency of optical energy injection into the localized modes, we define the focusing energy ratio $\Phi$, i.e., the ratio of the focus intensity $I_{\text{spot}}$ to the total transmitted intensity $I_{\text{tot}}$~\cite{leonetti2014light}
\begin{equation}
\Phi = \frac{I_{\text{spot}}}{I_{\text{tot}}}.
\label{eq:Phi_def}
\end{equation}
From a macroscopic physical picture, when the optical field couples into a transmission channel with a shorter localization length, the proportion $\Phi$ of the focused spot energy within the entire transmitted field becomes larger. Given that the refractive index is uniform along the propagation direction, we can reasonably assume that the transport efficiencies of different transmission channels remain a constant. It is not hard to infer that $\eta$ varies linearly with $\Phi$.
Next, we demonstrate these unique features of controlled Anderson localization.

Our experimental setup is depicted in Fig.~\ref{Figure_2}. A vertically polarized laser beam with a wavelength of 532 nm passes through a beam expanding system composed of lenses $L_1$ and $L_2$, and is then incident on a pure-amplitude spatial light modulator (SLM). We utilize a 4-f filtering system formed by $L_3$ and $L_4$ to achieve pure-phase control on the wavefront through the holographic method~\cite{hong2024robust}. The objective lens $O_1$ ($50\times$) projects the shaped wavefront onto the front surface of the photonic lattice, while a second objective lens $O_2$ ($40\times$) images the scattered light from the rear surface onto a camera. A horizontal polarizer is placed in front of the camera to block residual direct light.

In our experiment, the sample is a self-assembled two-dimensional ZnO lattice (Yiri DP450-300S-50000), as shown in Fig.~\ref{Figure_2}(a). This lattice consists of a periodic array of air holes with an average lattice constant of 450 nm, an average hole diameter of 300 nm, and a sample thickness of 50 $\mu$m. Due to the inherent disorder introduced during the fabrication process, the size, shape, and position of the holes exhibit random variations. Consequently, the transmitted field forms a typical speckle pattern, shown in Fig.~\ref{Figure_2}(b).

\begin{figure}[b]
	\centering
	\includegraphics[width=0.48\textwidth]{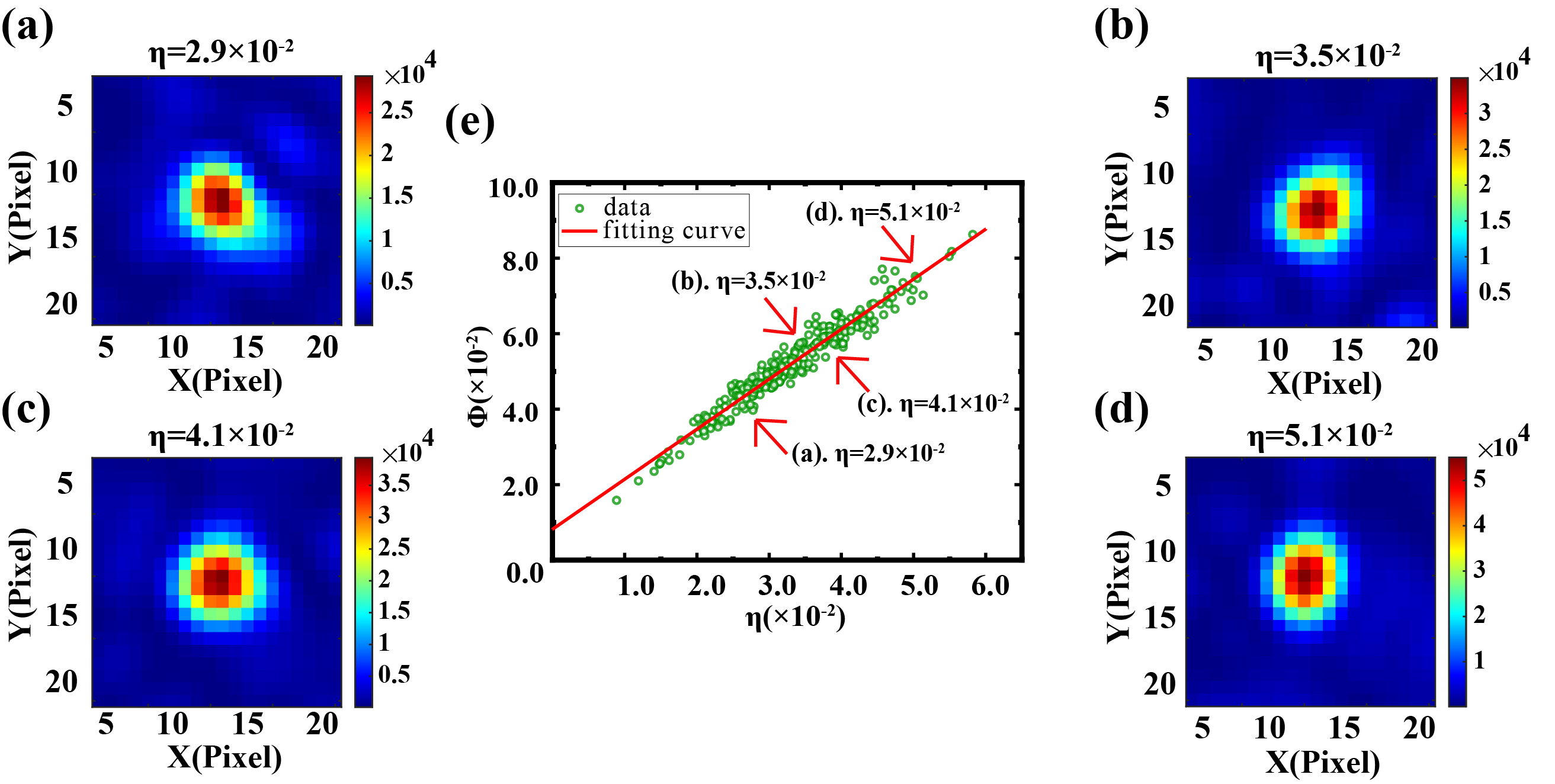}
	\caption{(a-d) Focusing patterns obtained under different focusing accuracies $\eta$. (e) Dependence of the relative ratio $\Phi$ of the focused energy on the focusing fidelity $\eta$.\label{Figure_3}}
\end{figure}

To achieve single-speckle focusing via wavefront shaping, we measure the transmission matrix $\mathbf{T}$ using a coaxial interferometry method~\cite{popoff2010measuring}. In experiment, the SLM is divided into $16 \times 16$ segments to control the incident wavefront, while a CCD camera simultaneously records the intensity at $16 \times 16$ positions. Consequently, the obtained transmission matrix $\mathbf{T}$ has $256 \times 256$ elements. With retrieved $\mathbf{T}$, we successfully realize focusing at arbitrarily specified positions, as shown in Figs.~\ref{Figure_3}(a-d).
Our statistical analysis over all results reveals the dependence of $\eta$ on $\Phi$, as plotted in Fig.~\ref{Figure_3}(e). A pronounced linear relationship between $\eta$ and $\Phi$ is clearly observed, which is fully consistent with our prediction.

\begin{figure}[t]
	\centering
	\includegraphics[width=0.48\textwidth]{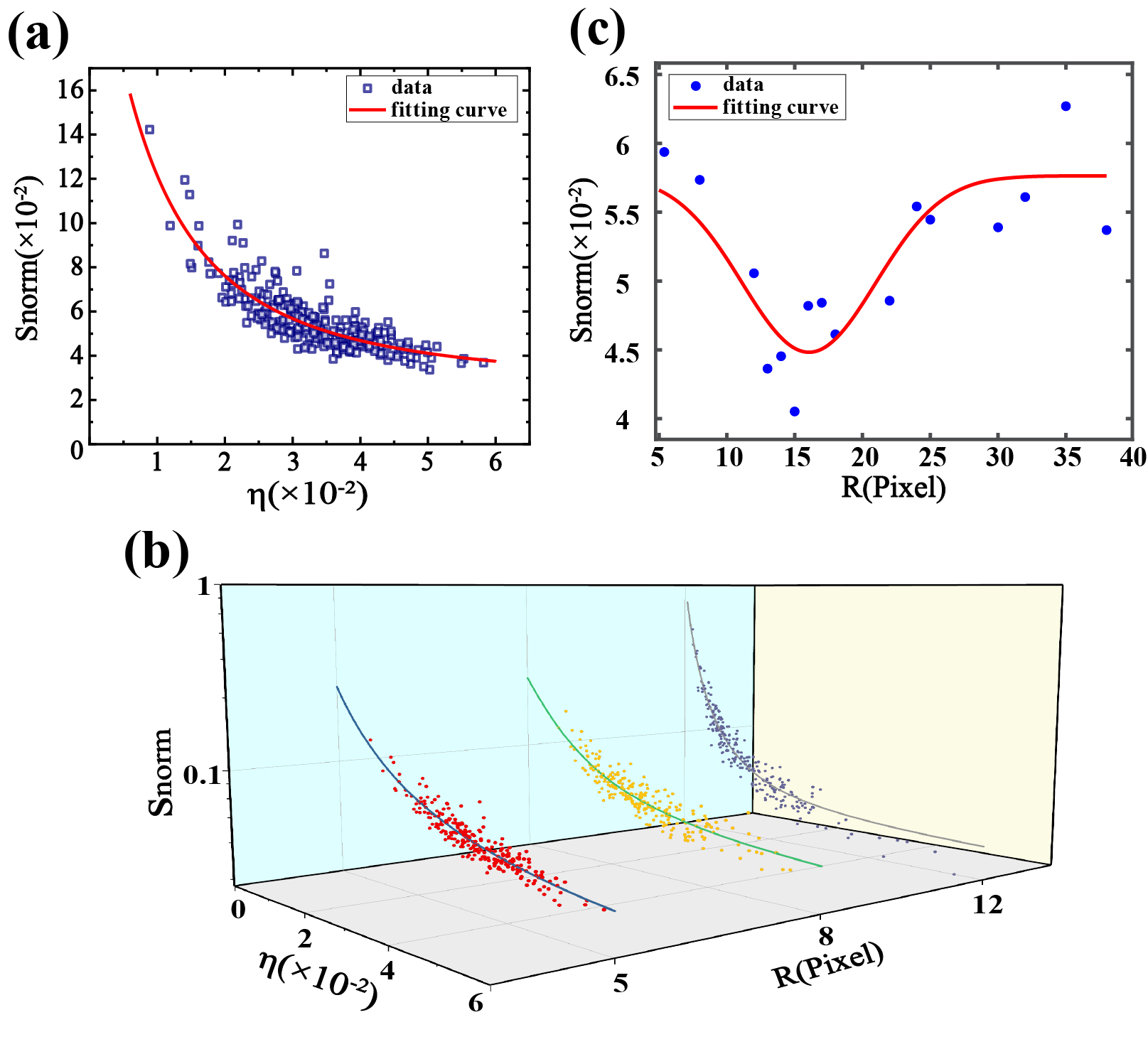}
	\caption{(a) Variation of the normalized speckle area $S_{\mathrm{norm}}$ with the focusing fidelity $\eta$. (b) Dependence of $S_{\mathrm{norm}}$ on $\eta$ with different focusing sizes $R$. The fitting curve is a Lorentzian function.	(c) Variation of $S_{\mathrm{norm}}$ with the focusing size $R$ at a fixed focusing fidelity $\eta = 0.029$, manifesting the critical coupling effect.	\label{Figure_4}}
\end{figure}

Next, we analyze the localization features of the scattered light.
To this end, we calculate the effective area of the scattered speckle field at the detection plane, defined as~\cite{choi2017self}
\begin{equation}
S_{\mathrm{loc}} = \frac{\iint_{\Omega} I(x,y) \, dx dy}{\max(I(x,y))},
\label{eq:S_loc_def}
\end{equation}
where $\Omega$ denotes the entire transmission plane and $I(x,y)$ is the intensity distribution. Subsequently, we introduce a dimensionless normalized area to quantify the concentration of the optimized speckle field
\begin{equation}
S_{\mathrm{norm}} = \frac{S_{\mathrm{loc}}}{S_{\mathrm{ref}}}.
\label{eq:S_norm_def}
\end{equation}
Here, $S_{\mathrm{ref}}$ is the area of scattered speckle field when a random wavefront is incident.

At single-spot focusing, the normalized speckle area $S_{\mathrm{norm}}$ for different values of $\eta$ is presented in Fig.~\ref{Figure_4}(a). One can see that a larger $\eta$ leads to a smaller $S_{\mathrm{norm}}$, agreeing with the conjecture that larger $\eta$ originates from hot spots of more compact localized modes. It is worth noting that the transmitted light is generally a superposition of multiple localized eigenmodes, implying that $S_{\mathrm{norm}}$ only describes the relative strength of localization.

Because a specific localized mode are unknown, we can not determine the input wavefront $\mathbf{E}_{\text{in}}$ required to excite a particular localized mode. To overcome this challenge, we define a variable focusing region on the transmission side, as illustrated in Fig.~\ref{Figure_1}(b).
Again, we employ wavefront shaping to enhance the transmitted light intensity within this target region, i.e. we optimize transmitted intensity $I_{\text{opt}}$ given by
\begin{equation}
I_{\text{opt}} = \frac{1}{S_{\text{opt}}} \iint_{\Omega_{\text{opt}}} I(x, y) \, dx \, dy.
\label{eq:I_opt_def}
\end{equation}
Here $\Omega_{\text{opt}}$ denotes the variable focusing region, and $S_{\text{opt}}$ is its area. 
It is straightforward to infer that as the size of the focal region increases and approaches the characteristic size of the minimum localized mode, the coupling efficiency of the light field into this mode improves, leading to a gradual decrease in the normalized area $S_{\text{norm}}$ of transmitted field. Conversely, when the focal region exceeds this intrinsic localization size, an increasing number of weakly localized modes are excited, resulting in a progressive increase in the normalized area $S_{\text{norm}}$ of transmitted field.
Hence, we predict that a critical coupling condition is achieved when the focusing region size matches the characteristic size of the minimum localized mode. At this point, the minimum localized mode is optimally excited, resulting in the smallest $S_{\text{norm}}$.

Next, we conducted experimental verification of the above prediction. We performed wavefront shaping for different focusing region. In experiment, the focusing region is a square with characteristic length $R$ as illustrated in Fig.~\ref{Figure_1}(b).
The obtained normalized area $S_{\text{norm}}$ consistently exhibits a decrease as $\eta$ increases as shown in Fig.~\ref{Figure_4}(b). However, the decreasing rate changes by altering the focusing size $R$.

\begin{figure}[b]
	\centering
	\includegraphics[width=0.5\textwidth]{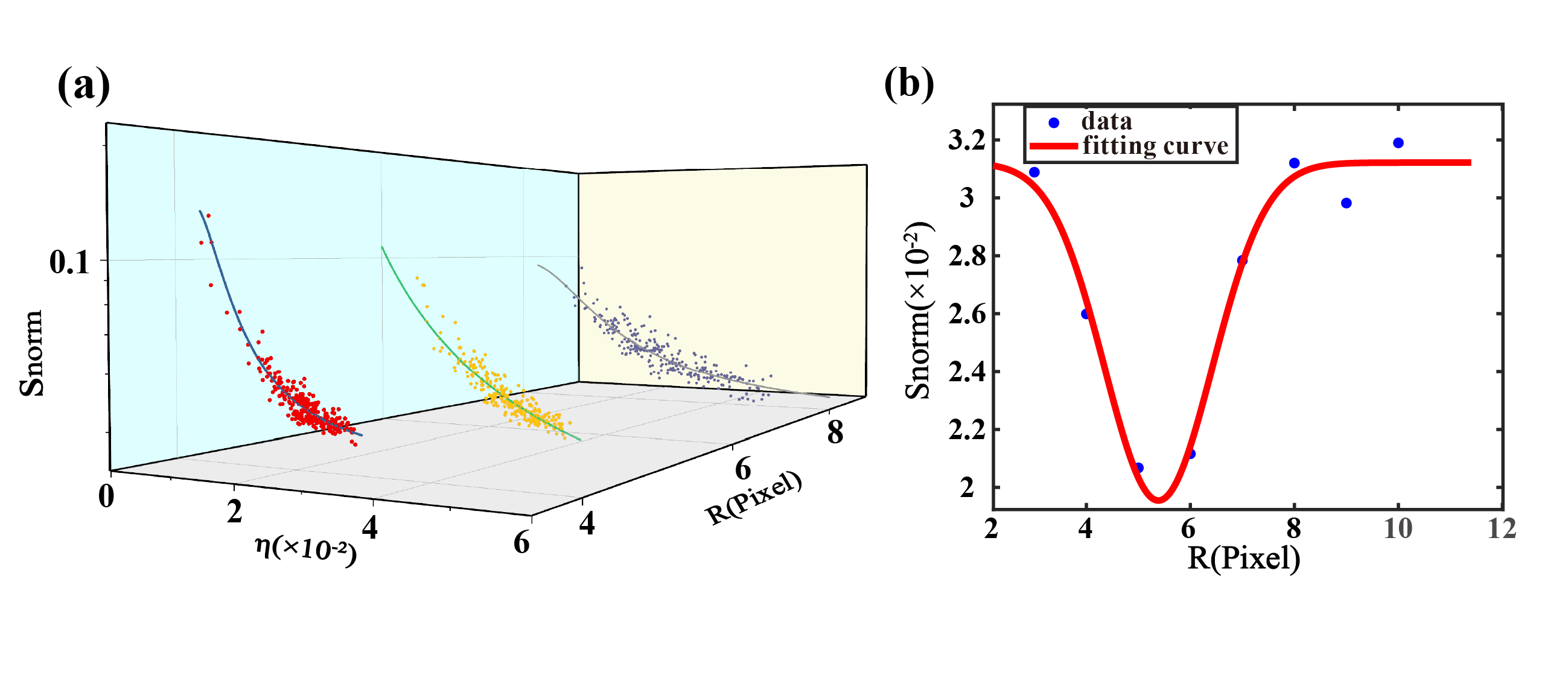}
	\caption{Experimental results for the second sample.
		(a) Dependence of the normalized speckle area $S_{\mathrm{norm}}$ on the focusing fidelity $\eta$ for different focusing sizes $R$.
		(b) The critical coupling effect.\label{Figure_5}}
\end{figure}

To quantitatively describe this variation, we further analyze the dependence of the localization length on the focusing radius at a fixed $\eta$ ($=0.029$). The results are shown in Fig.~\ref{Figure_4}(c). It can be seen that as the focusing size $R$ increases, the normalized speckle area $S_{\text{norm}}$ first decreases to reach a minimum, and subsequently increases before gradually saturating. This trend agrees well with our prediction, i.e. a smallest speckle area reaches when the focusing size is spatially matched with the minimum localized mode. We fit the data points in Fig.~\ref{Figure_4}(c) using the function $g(x) = a + b e^{-(x-c)^{2}/d^{2}}$, which yields a fitting parameter $c = 16.4$. Build on the critical coupling effect, the minimum localization size is derived to be 16.4 pixel.

To further validate this approach, we measured another sample (Yiri DP450-390S-50000) with different structure parameters. This sample shares the same average lattice period and thickness as the first one, but the average diameter of the air holes is increased to 390 nm. For different focusing sizes $R$, the normalized speckle area $S_{\text{norm}}$ decreases with increasing $\eta$, while the line shape exhibits significant variations as shown in Fig.~\ref{Figure_5}(a). Similarly, by fixing $\eta$ at a relatively high value $0.029$, we obtained $S_{\text{norm}}$ with respect to the focusing size $R$. The results are shown in Fig.~\ref{Figure_5}(b). Again, we observe the critical coupling effect, where the dependence of $S_{\text{norm}}$ on $R$ exhibits an obvious dip. By fitting the data points with the function $g(x) = a + b e^{-(x-c)^{2}/d^{2}}$, the minimum localization size is determined to be 5.4 pixel, which is only 32.9\% of that of the first sample.
Clearly, increasing the size of air hole leads to stronger localization in the self-assembled lattices.

In summary, we propose and demonstrate a scheme for determining the characteristic size of the minimum localized mode. 
By utilizing out-of-plane excitation integrated with wavefront shaping, we achieved high-efficiency coupling of incident light into the localized modes. 
By employing variable-size focusing, a critical coupling condition is found, which enables accurate determination of the intrinsic minimum localization size in disordered photonic lattices.
Experimentally, we report the observation of the critical coupling phenomenon, and use it to determine the minimum localization length in two different disordered lattices, providing a robust validation of the proposed approach.
This work not only provides a powerful tool for characterizing localization features in complex systems but also offers a pathway for efficiently exciting the localized modes.
Future research could explore selective excitation of localized modes in disordered systems via wavefront shaping, achieving active control of coupling and energy transport among multiple localized channels. This holds significant potential for applications in integrated optoelectronics, such as reconfigurable optical interconnects and logic units.

\vspace{12pt}
\begin{acknowledgments}
The authors wish to thank the support from National Natural Science Foundation of China (12574320); Shandong Provincial Natural Science Foundation (ZR2024QA225); the Open Project Funding of the Ministry of Education Key Laboratory of Weak-Light Nonlinear Photonics (OS26-1); Program for Innovative Research Team in Anqing Normal University.
\end{acknowledgments}

\bibliography{reference}

\end{document}